# The Need for Laboratory Work to Aid in The Understanding of Exoplanetary Atmospheres

Jonathan J. Fortney[a], Tyler D. Robinson[a], Shawn Domagal-Goldman[b], David Skålid Amundsen[c], Matteo Brogi[d], Mark Claire[e], David Crisp[f], Eric Hebrard[b], Hiroshi Imanaka[g,h], Remco de Kok[i], Mark S. Marley[h], Dillon Teal[a], Travis Barman[j], Peter Bernath[k], Adam Burrows[l], David Charbonneau[m], Richard S. Freedman[g,h], Dawn Gelino[n], Christiane Helling[e], Kevin Heng[o], Adam G. Jensen[p], Stephen Kane[q], Eliza M.-R. Kempton[r], Ravi Kumar Kopparapu[s], Nikole K. Lewis[t], Mercedes Lopez-Morales[u], James Lyons[v], Wladimir Lyra[w], Victoria Meadows[x], Julianne Moses[y], Raymond Pierrehumbert[z], Olivia Venot[zz], Sharon X. Wang[s], Jason T. Wright[s]

[a]University of California, Santa Cruz, [b]NASA Goddard Space Flight Center, [c]Columbia University and NASA Goddard Institute for Space Studies, [d]University of Colorado at Boulder, [e]University of St Andrews, [f]NASA JPL, [g]SETI Institute [h]NASA Ames Research Center, [i]Leiden Observatory, [j]University of Arizona, [k]Old Dominion University, [l]Princeton University [m]Harvard University, [n]NASA Exoplanet Science Institute, Caltech, [o]University of Bern, [p]University of Nebraska at Kearney, [q]San Francisco State University, [r]Grinnell College, [s]Penn State University, [t]Space Telescope Science Institute, [u]Harvard-Smithsonian Center for Astrophysics, [v]Arizona State University, [w]California State University Northridge and NASA JPL, [x]University of Washington, [y]Space Science Institute, [z]Oxford University, [aa]University of Leuven

**Abstract**

Advancements in our understanding of exoplanetary atmospheres, from massive gas giants down to rocky worlds, depend on the constructive challenges that occur between observations and models. Both from the ground and from space, we are now on a clear trajectory for improvements in exoplanet observations that will revolutionize our ability to characterize the atmospheric structure, composition, and circulation of these worlds. These improvements stem from significant investments by numerous funding agencies in new missions and facilities, such as the *James Webb Space Telescope* (*JWST*) and the several planned ground-based extremely large telescopes. However, while exoplanet science currently has a wide range of sophisticated models that can be applied to the tide of forthcoming observations, the trajectory for preparing these models for the upcoming observational challenges is unclear. Thus, our ability to maximize the insights gained from the next generation of ground- and space-based observatories is not certain. In many cases, uncertainties in a path towards model advancement stems from insufficiencies in the laboratory data that serve as critical inputs to atmospheric physical and chemical tools. This white paper was initiated within NASA's Nexus for Exoplanet System Science (NExSS), in collaboration with the larger exoplanet science community, to review the state of this science. We outline a number of areas where laboratory or *ab initio* investigations could fill critical gaps in our ability to model exoplanet atmospheric opacities, clouds, and chemistry. Specifically highlighted are needs for: (1) molecular opacity linelists with parameters for a diversity of broadening gases, (2) extended databases for collision-induced absorption and dimer opacities, (3) high spectral resolution opacity data for a variety of relevant molecular species, (4) laboratory studies of haze and condensate formation and optical properties, (5) significantly expanded databases of chemical reaction rates, and (6) measurements of gas photo-absorption cross sections at high temperatures. We hope that by meeting these needs, we can make the next two decades of exoplanet science as productive and insightful as the previous two decades.

**Introduction**

The next two decades will see unprecedented growth in the field of exoplanet atmospheres. The *Transiting Exoplanet Survey Satellite* (*TESS*)[1], scheduled to launch in 2017, will build on the *Kepler* mission's[2] legacy by detecting transiting exoplanets around the Sun's nearest neighbors. The *James Webb Space Telescope* (*JWST*)[3], which will launch one year after *TESS*, will use transit and secondary eclipse spectroscopy, direct imaging, and phase curve studies to provide glimpses into exoplanet atmospheres over a wide range of wavelengths and at a spectral resolution up to ~3,600[4]. Most excitingly, the *JWST* will have the capabilities to probe the murky boundary between rocky planets and gas giants, and may even investigate the best potentially habitable terrestrial exoplanet targets discovered by *TESS*[5]. Finally, the mid-2020s will see the launch of the *Wide Field Infrared Survey Telescope* (*WFIRST*)[6] and first light for several extremely large ground-based telescopes (ELTs)[7,8,9], breaking open the field of exoplanet characterization in reflected light at, in some cases, extremely high spectral resolution.

Nevertheless, even with our current generation of instruments and telescopes, exoplanets present us with many unsolved mysteries. For the lowest mass planets that we currently investigate, hazes of unknown composition and source commonly block our ability to study atmospheric chemistry[10,11]. Observations of certain hot Jupiters demonstrate what may be cloud effects on the night sides of these planets that current models struggle to explain[12]. Finally, the first results from the Gemini Planet Imager (GPI)[13] and Spectro-Polarimetric High-contrast Exoplanet REsearch (SPHERE)[14] instrument direct imaging surveys of hot, young giant exoplanets are revealing gaps in our understanding of giant planet evolution and thermal structure[15,16]. While the incredible exoplanet characterization power of *JWST*, *WFIRST*, and ELTs will bring higher signal-to-noise data and incredible spectral resolution to bear on existing exoplanet mysteries, it is certain that these observatories will present even greater challenges to our understanding of planetary atmospheric physics and chemistry.

Of course, the ultimate motivation for all of the aforementioned investments in tools to study exoplanet atmospheres is the overarching goal to characterize the nearest habitable exoplanets for signs of life. For example, understanding the best techniques to characterize the atmospheres of hazy super Earths prepares us for our eventual studies of habitable Earth-like planets, which will almost certainly be cloudy and may even possess a photochemical haze layer[17]. Thus, it is important that we maximize the exoplanet lessons that we can learn from missions and instruments in the next decade. As has already been demonstrated in some cases, though, it is likely that our biggest impediments to understanding exoplanets in the future will not be the quality of observations, but will instead be the quality of our models for atmospheric processes and their associated input data.

Ever since the rise of the field of planetary science, it has been clear that to model, interpret, and understand planetary atmospheres that we must depend on laboratory expertise to deliver necessary physical and chemical inputs and constraints. While Earth-like conditions and processes tend to be emphasized, Solar System studies have shown us a need to understand atmospheres both far hotter and colder than our own, with compositions far outside our everyday experience[18]. Exoplanets are now further broadening our point of view, and have demonstrated a need to understand the role of seemingly exotic regimes of atmospheres, including clouds of rock dust and iron droplets, the potential for sulfur-derived hazes, and high-temperature opacities for conditions more similar to cool star atmospheres than to any Solar System world[19,20]. The picture is further complicated by the complexities (and uncertainties) of atmospheric escape as well as outgassing from planetary interiors.

Given the huge expansion in exoplanet observing capabilities that will occur over the next decade, now is the ideal time to probe for gaps in our understanding of the most fundamental processes at work in planetary atmospheres—gas opacities, thermo- and photo-chemical reactions, and cloud and haze chemistry and properties. By identifying and rectifying these gaps, we can ensure our ability to gain maximum insight from missions like *JWST* and *WFIRST*, and best prepare for the successors to these missions. This document began as a collaborative project within NASA's Nexus for Exoplanet System Science (NExSS) and grew to include many investigators outside of NExSS. It is a discussion of currently-known gaps in our understanding of key physical and chemical quantities relevant to the study of exoplanetary atmospheres. Included in this discussion are proposals for how these gaps can be filled. In many cases, given the lack of data, it is difficult to quantify the extent to which new laboratory results will impact existing models. Nevertheless, given that most databases tend to emphasize relatively clement Earth-like conditions, we can expect significant, and enlightening, changes as we move to the extremely exotic atmospheres of worlds around other stars.

**Pressure Induced Line Broadening Parameters**

Pressure induced line broadening parameters are crucial to the computation of atmospheric opacities, which in turn determine planetary spectra and atmospheric heat budgets[21]. By perturbing radiatively active molecules through collisions and dipole-dipole interactions, gases with large partial pressures induce broadening of spectral lines, and may in some cases cause line profiles to become significantly non-Voigt-like. For Earth's atmosphere, these broadening parameters are readily available in the HITRAN line database[22]. Pressure-induced line widths induced by both air (foreign broadening) and the molecule itself (self broadening) are given for each line. Extrapolation was necessary when constructing the high temperature version of HITRAN, HITEMP[23], where air- and self-broadened widths were extrapolated to both higher quantum numbers and higher temperatures, although some work exists to improve the air- and self-broadened widths for water at higher temperatures[24].

However, for atmospheres dominated by $H_2$ and He, no database with pressure broadening parameters exists, but some limited experimental and theoretical data are available[25,26,27,28]. These data are extremely sparse, as they are only available for relatively few lines, and only valid for a very limited temperature and pressure range. Consequently extrapolation of these parameters by as much as an order of magnitude outside of their intended region of validity becomes necessary. Similarly, atmospheres dominated by $CO_2$, such as those of Venus, Mars and the Archean Earth, require $CO_2$ pressure-broadening parameters for the radiatively active species. Due to lack of availability of such data, air-broadened widths are usually used[29,30], although some limited data for $H_2O$ lines is available[31,32].

In addition to lab measurements, theoretical *ab initio* calculations can help improve pressure-broadening parameters, particularly as measurements for a large number of lines and a large temperature range may be difficult to obtain. Such calculations require lab measurements for tuning purposes. However, these can be performed at room temperature and pressure and for a smaller number of lines[33,34,35].

Below is an overview of the data required. A common database with easily accessible data would provide for broad straightforward use by the community.

- $N_2$ and/or $O_2$ dominated atmospheres:
    - Radiatively active species (broadeners):
        - $H_2O$ ($N_2$, $O_2$, $H_2O$)
        - $CO_2$ ($N_2$, $O_2$, $CO_2$)
        - $CH_4$ ($N_2$, $O_2$, $CH_4$)
        - $O_3$ ($N_2$, $O_2$)
    - Temperature range: 70 K to 500 K
    - Pressure range: Up to ~10 bar
- $CO_2$ dominated atmospheres:
    - Radiatively active species (broadeners):
        - $H_2O$ ($CO_2$, $H_2O$)
        - $CO_2$ ($CO_2$)
        - $CH_4$ ($CO_2$, $CH_4$)
    - Temperature range: 70 K to 2000 K
    - Pressure range: Up to ~100 bar
- $H_2O$ dominated atmospheres:
    - Radiatively active species (broadeners):
        - $H_2O$ ($CO_2$, $H_2O$)
        - $CO_2$ ($CO_2$, $H_2O$)
        - $CH_4$ ($H_2O$, $CH_4$)
    - Temperature range: 70 K to 2000 K
    - Pressure range: Up to ~100 bar
- $H_2$ and He dominated atmospheres:
    - Radiatively active species:
        - $H_2O$, CO, $CO_2$, $CH_4$, $NH_3$, Na, K, Li, Rb, Cs, TiO, VO, HCN, $C_2H_2$, $H_2S$, $PH_3$
        - Broadeners: $H_2$ and He
    - Temperature range: 70 K to 3000 K
    - Pressure range: Up to ~100 bar
- Vaporized terrestrial atmospheres[36]
    - Radiatively active species:
        - $CO_2$, $H_2O$, $SO_2$, HCl, HF, OH, CO, SiO, KOH, KCl
        - Broadeners: $CO_2$ and $H_2O$
    - Temperature range: 700 K to 4000 K
    - Pressure range: Up to ~100 bar

**Lab Spectroscopy of Continuum Absorption**

Pressure-induced continuum absorption in planetary and brown dwarf atmospheres comes in two forms. First, frequent collisions in dense atmospheric regions can induce transitory dipole moments in key molecules, thereby permitting transitions that are ordinarily forbidden. This so-called "collision-induced absorption" (CIA) is generally smooth in wavelength, as the colliding molecule carries away non-quantized translational kinetic energy. Second, supramolecules (e.g., dimers, trimers, etc.), which are in general short-lived, can reach appreciable steady-state concentrations due to recurrent collisions in high-density regions of a planetary atmosphere. These supramolecules tend to create broad, smoothly-varying absorption features because the uncertainty principle indicates that very short-lived phenomena (like the formation of a collisional dimer) will have a relatively large spread in energy states.

The importance of pressure-induced absorption (both CIA and dimer absorption) in determining atmospheric thermal structure and its role in sculpting planetary and brown dwarf spectra is well-known. Collision-induced absorption from $H_2$, He, and H are of central importance for understanding the thermal structure and spectra of gas giants[37], and also provides the continuum level at many wavelengths for cool brown dwarfs[38,39,26]. Additionally, $H_2$–$H_2$ CIA could play an important role in determining the greenhouse effect (and, thus, extending the habitable zone) for super-Earths with hydrogen-enriched atmospheres[40]. Various combinations of $N_2$, $CH_4$, and $H_2$ CIA provide the primary greenhouse effect on Titan, and $CO_2$–$CO_2$ pressure-induced absorption is a key opacity source for Venus (and likely played an important role in any $CO_2$-rich atmospheres in the early Solar System, including the early Earth). Finally, molecular oxygen and nitrogen dimer absorption have been detected in Earth's transmission[41] and disk-integrated spectra[42], and could be used as a pressure indicator for exoplanets[43].

Compilations of large amounts of CIA data are available[44], and contain results from laboratory and modeling studies. Most results tend to emphasize $N_2$ and $O_2$, as these are the dominant molecules in Earth's atmosphere, and the studied temperatures ranges are largely focused on our planet. Results for $H_2$ also exist[45], and have been successfully adapted to the range of conditions from Solar System gas giants to brown dwarfs[24].

Despite the importance of pressure-induced absorption for shaping our understanding of planetary atmospheres and habitability, several key holes still exist in our knowledge. Some of these gaps are related to a simple lack of opacity data (or modeled opacities). For example, recent work on the early Martian atmosphere was forced to use $N_2$-$H_2$ CIA in the place of $CO_2$-$H_2$ CIA[46], as no data for the latter currently exists. Additionally, in a problem relevant to climate on modern Earth as well as the runaway greenhouse, there is uncertainty over how much of the so-called "water vapor continuum" is due to water vapor collisional dimers versus a super-Lorentzian adjustment to water vapor lines[47], which contributes to modest discrepancies in models of the inner edge of the Habitable Zone[48,49]. Finally, in particularly dense planetary atmospheres (e.g., Venus), it is likely that trimer (or larger) supramolecules could provide continuum opacity, which has been largely unstudied.

It is clear that future lines of laboratory (or model) investigation that shed light on $CO_2$-$H_2$ CIA and water vapor dimer absorption should be encouraged. Databases should be unified and standardized (as in, e.g., Richard et al. 2012). Additionally, gaps in our existing databases for dimer absorption or for absorption by key bulk atmospheric constituents ($H_2$, $N_2$, $CO_2$, $O_2$) generated by collisions with relevant background gases ($H_2$, $N_2$, $CO_2$, $O_2$, $H_2O$, $CH_4$, $CO$, $NH_3$) should be filled in, and temperature ranges for these databases should be extended to conditions appropriate for the broad range of environments encountered in planetary astrophysics (i.e, 50–3,000 K). Such data would have an immediate impact on our understanding of planetary habitability and Earth's climate. In practice, mechanisms could be put in place wherein new CIA and/or dimer absorption data can be requested by planetary atmospheric scientists. Either new laboratories (or funding calls) would need to be set up to respond to such requests. A level of peer review would help rank the data requests that are most likely to yield results of the highest importance for studying the diversity of atmospheres that exoplanet scientists are likely to encounter in the coming decades.

**Molecular Opacity Data at High Spectral Resolution**

In the last few years, ground-based high-resolution spectroscopy has been increasingly used to study the properties of exoplanet atmospheres in the near-infrared[50,51,52,53,54,55,56]. This method uses the cross correlation between the data and a range of template spectra to combine and enhance the signal of the many molecular lines in the exoplanet spectrum. However, the computation of template spectra relies on the knowledge of the position and strength of spectral lines of the most relevant molecular species, which for giant planets orbiting close to their parent star (hot Jupiters) are $H_2O$, $CO$, $CH_4$, and $CO_2$ in the near infrared[57,58], and potentially TiO or VO in the optical[59,60]. Any inaccuracies in the line lists may lead to partially incorrect templates, which might have a serious impact on the measured cross-correlation functions (CCFs). Whereas slightly incorrect cross sections or a few missing weak lines would not affect the CCF significantly, entirely missing bands, systematic shifts or random inaccuracies in the positioning of the deepest lines would reduce the power in the peak CCF, in turn reducing the measured S/N. Furthermore, if measuring line broadening through the CCF is necessary[61], line positions must be accurate at a fraction of the FWHM of the spectrograph. For spectral resolutions of 100,000 as commonly achieved in most of the past studies, this means an accuracy better than 1 km s$^{-1}$ (or, e.g., 0.01 cm$^{-1}$ at 2.3 μm).

A primary issue is that in hot Jupiters (and other hot planets) millions of lines that are very weak at room temperature become much more pronounced. Many of the room temperature based lab measurements and databases (HITRAN, GEISA) are thus inadequate for high T. It is therefore crucial to improve and expand line lists for use with high-dispersion observations. Ideally, complete and reliable catalogs would be available by the end of the decade, when high-dispersion observations might offer the best chance for ground-based follow up for the most promising exoplanet candidates found by missions such as TESS.

Presently, CO and $H_2O$ have the most reliable line lists, as demonstrated by the long list of detections reported to date (see above), one of which was even used to infer planet rotation and atmospheric winds from the broadening of the CCF[61]. However, some of the spectral regions for $H_2O$ are known to be inaccurate, and revisions of the HITEMP[62] and ExoMol[63] line lists are underway.

The line lists for $CO_2$ significantly changed between HITEMP 1995[64] and HITEMP 2010[62]. Anomalous strong lines that are present in the old database at 2.3 μm and 3.2 μm disappeared in the revised catalog[65]. It is therefore advisable that future work revisit this molecule and address such discrepancies.[*]

Methane line lists underwent several revisions in the last years. The ExoMol line list[66] has been used extensively for computing brown dwarf spectra. Another recent catalog is that of Hargreaves et al. (2015)[67]. One common outcome of these recent line lists is that weak $CH_4$ lines differ drastically at low and high temperatures. Unfortunately, no detection of methane in an exoplanet spectrum has yet been reported at very high spectral resolution. However, methane has been detected in the atmospheres of directly imaged planets through moderate-[68] and low-resolution[14] spectroscopy.

---

[*] Improvement in the line list and line strengths for species at typical Earth atmosphere conditions in the HITRAN database is also crucial for the detection of exoplanets via precise radial velocities (RVs). Remaining limitations of our knowledge of Earth's atmospheric absorption lines is a major bottleneck for achieving 10 cm/s RV precision to detect Earth analogs or for achieving below 3 m/s in the NIR. As summarized in the conference proceeding for the Second Workshop on Extreme Precision Radial Velocities at Yale in July 2015 (Fischer et al. 2016, PASP submitted), the RV community recommended more laboratory and theoretical work to be done to refine the HITRAN database for better modeling of the telluric lines and removal of this source of spectral contamination (e.g., most prominently, $H_2O$ and $CH_4$ lines).

In the optical, TiO and VO are thought to be two potentially abundant high-altitude absorbers capable of producing inversion layers in the atmospheres of hot Jupiters (i.e., a region where temperature increases with altitude). Searching for these species in the transmission spectra of giant planets is therefore of great importance to understand one of the basic mechanisms regulating the energy budget in exoplanet atmospheres. However, as Hoeijmakers et al. (2015) demonstrated[69], current line lists for TiO are incomplete and inaccurate. The cross-correlation between a template model and an M-dwarf spectrum is in fact almost zero across large portions of the spectrum, and a visual comparison between the observed data and the models reveal a poor match in the line position.

Finally, one of the strengths of high-dispersion observations is the ability to measure relative molecular abundances. With minor species targeted, it will be also possible to distinguish between carbon-rich planets (i.e., planets in which all the C-bearing molecules are overabundant compared to the O-bearing compounds) from disequilibrium chemistry (i.e., cases in which just one or few of the species are out of equilibrium abundance). This type of study requires accurate line lists for a number of additional species, including but not limited to $NH_3$, $C_2H_6$, $C_2H_4$, $C_2H_2$, and HCN.

**Lab Experiments on Haze Formation**

Atmospheric aerosols play central roles in the dynamically, radiatively, and chemically coupled system of exoplanetary atmospheres. Constraining chemical compositions of a wide variety of exoplanet atmospheres is of primary importance to understanding their origins, diverse physico-chemical natures, and their habitability. However, some recent transit observations have demonstrated the possibility of a wide prevalence of haze/cloud layers at high altitudes, obscuring the detection of major atmospheric constituents. As a result, we must understand plausible formation mechanisms of haze/cloud particles and their optical properties to interpret the observational data. Furthermore, such laboratory investigations might provide insights into techniques for characterizing the pressure-temperature profiles, C/O ratios, and metallicities of exoplanetary atmospheres. Understanding of the formation chemistry, thermal stability, and optical properties of clouds and photochemical hazes in warm/hot exoplanetary atmospheres will be essential to interpreting future spectroscopic observations of these worlds.

Photochemical organic aerosols are ubiquitous among the atmospheres of the cold outer Solar System where significant $CH_4$ is present[70,71,72]. The dominant carbon-bearing gaseous species in exoplanetary atmospheres would be CO, $CH_4$, and $CO_2$, whose ratio depends on their metallicity, C/O ratio, temperature, and non-equilibrium processes. Several photochemical-thermal equilibrium models have explored the CHNO exoplanetary atmospheric chemistry in warm/hot exoplanets[71,72,73,74,75,76,77,78,79,80,81,82,83,84]. In heavily UV irradiated atmospheres of hot Jupiters, however, ion-molecule chemistry in the ionosphere, which plays crucial roles in generation of atmospheric aerosols on Titan[71,85,86,87], is not considered yet, aside from limited studies of atomic and simple molecular ions in hot Jupiter thermospheres[88] and stratospheres[89]. Dedicated laboratory investigations for various exoplanetary atmospheres would provide insight on the role of coupled ion and neutral chemistry in photochemical haze generation.

In the high temperature, low pressure, heavily UV irradiated conditions of some exoplanetary upper atmospheres, coupled chemistry of volatile elements (CHON) and other refractory elements, such as sulfur, phosphorus, alkali metals, and silicate/metal vapors, could generate dust particles, whose chemical compositions, structures, optical properties, and thermal stabilities have not been well investigated. Such systematic laboratory investigations (plausible haze compositions and their optical properties) is necessary for interpretation of the current and near-future observations of exoplanetary atmospheres. Studies under both early Earth and Titan-like conditions[90,91] reveal complex roles for O and S incorporation into haze particles, potentially altering their radiative properties[92]. The radiative imprint of hazes on early Earth's methane rich atmosphere persist in Earth's sulfur isotope record[93,94], providing a unique ground-truth opportunity to enhance modeling of hazy exoplanets.

• Further kinetic studies of chemical processes in gas-to-particle conversion and in heterogeneous gas-particle reaction processes, including the study of ion-molecule and UV photochemical driven growth.

• Measurements of vapor pressures of refractory materials. These data are critical to estimate formation of condensation clouds. However, there is general lack of laboratory measurements of accurate vapor pressures lower than $10^{-6}$ mbar. Heterogeneous condensation can occur as well; however, vapor pressures measurements of plausible gas mixtures are lacking.

• Measurements of particle growth and loss rates, chemical and thermal stabilities under plausible reactive environments for exoplanetary atmospheres.

**Studies of Refractory Condensate Clouds**

In addition to studies of stellar photon driven hazes, our understanding of refractory condensates that form in high temperature atmospheres (e.g., magnesium silicates, corundum, iron, perovskite) is sorely lacking, both for giant planets, and potentially for extremely strongly irradiated rocky planets. The condensation pathways, as a function of temperature, and the depletion of gaseous material due to condensation is a very active area of research, with diverse points of view with heritage in the solar system and stellar atmospheres literature[95]. Understanding the detailed process that lead to the condensation of cloud particles depends on the local gas temperature and pressure and is key to model the location at which such clouds may form. The equilibrium condensation sequence over a range of temperatures, pressures, and metallicities serves as a point of departure for understanding more realistic–and more complex–systems. Thus both the equilibrium condensation sequence and departures which follow different chemical pathways are worthy of study.

Such condensates are required to model young giant planets, such as the distant ones in the HR 8799[68] and 51 Eri systems[14], in addition to the close-in hot Jupiters. But yet there are virtually no lab data to inform such models. An example of an open question is to what extent does Fe enter silicate grains, although in a cooling gas the initial condensation of Fe occurs ~100 K before silicates. Fe is present at its vapor pressure while the silicate grains form and could be incorporated into these solids. Iron bearing silicate grains have very different optical properties than the pure Mg end member phases[96], yet most brown dwarf atmosphere models assume the pure magnesium forsterite and enstatite end members represent 100% of the olivines and pyroxenes present. Alternative pathways that emerge from kinetic cloud formation models[93] predict the formation of mixed particles that are made of silicates plus iron inclusion and these pathways likewise depend on sometimes sparse laboratory equilibrium data for a range of materials. While there have been lab studies of grain growth under solar nebula-like conditions, there is essentially no laboratory data under relevant conditions (e.g., ~1600 K and 1 bar atmosphere of $H_2$). Grains that form at lower temperatures, such as Cr, MnS, FeMgS, $Na_2S$, ZnS, and KCl, are also important for brown dwarf and exoplanet atmospheres[97,98] and have likewise been neglected.

Issues worthy of study include the extent to which these condensates are 'pure' or rather heterogenous "dirty" mixtures of multiple species[99,100,95], the shape of the grains, growth rates, and parameters relevant to microphysics particle growth calculations, such as cohesion properties. Again, studies conducted at relevant temperatures (500-2500 K for the major grain species) and pressures (10 bars to 10 mbar) would be most suited to improve the current state of atmospheric modeling. A cogent summary of the relevant species can be found in Lodders & Fegley (2006)[101].

Kinetic theories, which follow grain formation around seed particles, likewise require an understanding of the material properties for seed candidates. These include the surface growth/evaporation processes relevant to candidate seed formation (e.g., Table 1 in (98)). On Earth, condensation seeds are swapped up by winds from the crust into the atmosphere, but this will not always be possible for exoplanets. Required lab data therefore include:

- Cluster data that allow the inferences of chemical pathways from the gas-phase to the formation of increasingly more stable clusters.
- Surface reaction rates for each of the elementary reaction participating in the growth of an existing grain surface.
- A complete description of cloud formation processes would then be applicable to a broad range of conditions, including gasses of varying metallicities.

**Optical Properties of Particles**

Particles appear to be ubiquitous in atmosphere-bearing worlds. They are present in all the other solar system planets with stable atmospheres, including Earth, and have been invoked to explain the preliminary spectra of a number of exoplanets[102,103,104]. Among current work, the 6 Earth-mass planets GJ1214b is a particularly noteworthy example, where some kind of cloud opacity appears to be completely obscuring gaseous absorption features[10,105,98]. Particles also have a profound effect on the radiative transfer of a planet's atmosphere[106], which in turn has consequences for the planet's climate and for its observed spectrum.

Unfortunately, particle data are sparse and nowhere near diverse enough to account for the wide variety of atmospheric particles we anticipate observing on exoplanets. Much of the existing data sets are related to Earth-specific particles, or those observed in other solar-system worlds. And even in those cases, the data are often specific to the atmospheric conditions we have already encountered.

For example, consider organic aerosols (also referred to as "tholins" or as "hydrocarbon aerosols"). We have measured optical properties for these particles as they exist on modern-day Earth and modern-day Titan[107]. However, despite significant modeling work on the impacts of such particles on ancient Earth[108], we have no measurements that would be appropriate to the atmospheric composition of Earth at that time, which we think was less oxidizing than modern Earth yet much more oxidizing than modern Titan. This leaves us with no modern planet within which to make measurements of the aerosol formation pathways or the optical properties of the resulting aerosol. This is critical, because we know that the composition of organic aerosols can affect their optical properties, and that those that would have formed in the atmosphere of early Earth would have had different optical properties than those on Titan[109]. And this is the case for particles that have similar types of compositions (organic aerosols), just with different specific molecular constituents.

This goes well beyond organic aerosols, as well. Even when we limit our thinking to Earth history, we have other examples of aerosols whose optical properties are poorly understood, or only known within different atmospheric or chemical contexts. Sulfur is known on modern Earth to form sulfate aerosols, but in less oxidizing atmospheres similar to Archean Earth can also form particulate elemental sulfur or polysulfur (S4 or S8). The atmosphere of hot sulfur rich Io shows similar diverse chemistry. The properties of these particles is not nearly as well-studied as organic aerosols. Thinking further back in Earth history, relevant physical conditions and atmospheric compositions appropriate for study vary from the post-giant impact Earth atmosphere[108] through the Hadean. For gas giants this could include particles created from more than just methane-derived photolysis. In warmer giant planet atmospheres, $CH_4$ can be found mixed with CO, $NH_3$, $N_2$, $H_2S$, $PH_3$, such that condensed particulates could readily incorporate a wide range of compositions.

Three things are needed. (1) First, measurements of the optical properties of a wider diversity of particles in a greater range of atmospheric conditions. This includes measurements of organic aerosols in a wider variety of chemical contexts (especially different O abundances), measurements of elemental sulfur and polysulfur particles, measurements of particles for post-impact atmospheres, those on gas giants in our solar system, and finally those predicted for hot Jupiters. These measurements of optical properties should be made over a broad range of wavelengths, from the ultraviolet to the mid-infrared, to ensure their effects on incoming starlight for a variety of stars is accounted for, as well as any greenhouse forcings they impart. These measurements may also enable characterization of the aerosols themselves, if the aerosol being measured has specific features associated with its composition. (2) Second, we need a database in which to store these data so that they are accessible by the modeling and observing community that would utilize them. (3) Finally, these measurements must be made in the context of better understanding of the atmospheric formation pathways for these species (see section on Lab Studies of Haze Formation).

**Reaction Rate Constants**

The expected diversity of exoplanetary atmospheres spans a broad range of physical and chemical conditions. Reaction rates must therefore be known at temperatures ranging from ~30 K to above 3000 K, and because the deep atmospheric layers are chemically mixed with the layers probed by spectroscopic observations, at pressures up to about 100 bars.

Reaction rate constants that are important in Earth's current atmosphere are generally known with accuracy. There is however quite a large parameter space of temperature, pressure, and composition conditions that have not been thoroughly studied experimentally. Broadly speaking, these encompass any reactions that do not occur under modern Earth conditions. This bias limits the study of Solar system planets, as well as theoretical studies into the early Earth (when the atmosphere was reducing rather than oxidizing) as well as exoplanets.

Chemical models used to study the chemical composition of planetary atmospheres are based on a network of chemical reactions and associated rate coefficients. These reactions and rate coefficients are partially compiled from data found in the literature, when available. The KInetic Database for Astrochemistry (KIDA, http://kida.obs.u-bordeaux1.fr) is a recent chemical database that consists of reactions with rate coefficients and uncertainties that are evaluated to the best extent possible. Submissions of rate coefficients measured and calculated in the laboratory are studied by experts before inclusion into the database. A subset of the reactions in the database (kida.uva) is regularly provided as a network for the simulation of the chemistry of dense interstellar clouds and planetary atmospheres with temperatures between 10 K and 300 K. Also, fortunately, hot exoplanets conditions are found in typical combustion experiments and explosion engines and we benefit from decades of research on the chemical kinetics needed for exoplanetary atmospheres. Thus, chemical schemes validated as a whole against experiments have been developed[79,111]. The range of validation being similar to the conditions found in hot exoplanetary atmospheres (300-2500 K), these kind of schemes are quite reliable (and openly available as well on the KIDA web site).

To further our understanding of the link between chemistry and aerosols in exoplanetary atmospheres, it is important to expand the gas and condensed-phase reaction mechanisms to include the chemistry of species other than C, H, O, N. There is however a lack of reliable kinetic data, and in some cases thermodynamic data, in the scientific literature for molecules containing more exotic elements such S, P, Si, Mg, Na, K, Ca, Al, and Ti. Even the reaction rates and mechanisms for nitrogen species under hot giant-exoplanet conditions are poorly understood. For instance, it has been shown that the use of different sub-nitrogen networks (all validated and used in the combustion field) in exoplanetary models can lead to different predictions for the atmospheric composition[79]. There is also a lack of thermodynamic and kinetic data for heavier organic molecules and ions that might form in exoplanet atmosphere.

The disparate fields of air pollution, volcanic eruption chemistry, as well as studies of Titan's atmosphere, has enabled some study of the reactions governing organic haze formation, but again, the parameter space of these reactions needs to be expanded to atmospheres with different redox conditions. For the early Earth, incorporation of oxygen, sulfur, and nitrogen species into gaseous precursors and haze particles has not been well studied experimentally. For $H_2$-rich atmospheres, adding comprehensive sulfur and phosphorus reaction mechanisms are likely the next step, and can be guided by Solar-System atmospheric studies, astrophysical studies, and/or industrial applications. The gas-phase kinetics of alkali species is poorly known, other than for a few reactions that are important for meteoric or volcanic debris in planetary atmospheres, or chemistry in the interstellar medium, protoplanetary disks, molecular clouds, and other astrophysical environments.

Chemistry and dynamics are often tightly entangled in planetary atmospheres. For hot Jupiters, for instance, some molecules ($CH_4$, $NH_3$, $CO_2$ and HCN) can exhibit very different abundance profiles when including horizontal and/or vertical mixing[112,82]. Taking chemical processes into account within future General Circulation Models (GCMs) is critical, not only for guiding the future characterization, but also for the general understanding of exoplanet atmospheres and the interpretation of available observations. Along this same vein, the reactions controlling the transport-induced quenching of key molecular groups such as CO-$CH_4$-$H_2O$ and $NH_3$-$N_2$ on brown dwarfs and extrasolar giant planets are poorly understood due to of a lack of accurate rate-coefficient information at high temperatures and high pressures in $H_2$-rich environments. This lack of data strongly affects abundance predictions of spectrally-active molecules like CO, $CH_4$, $NH_3$, and $H_2O$ on exoplanets and sub-stellar objects. As discussed by Moses (2014)[113], particularly critical are reactions that convert between C-O bonded organics (e.g., $CH_3OH$, $CH_2OH$, $H_2CO$) and C-H and O-H bonded species (e.g., $CH_3$, $CH_4$, OH, $H_2O$), and reactions that convert between N-N bonded species (e.g., $N_2$, $N_2H_x$) and N-H bonded species (NH, $NH_2$, $NH_3$).

Rate-coefficient information is also needed for kinetic reactions dealing with the fate of the hydrocarbon radicals $C_3H_2$ and $C_3H_3$, which form readily on hot Jupiters from the insertion of C and CH into $C_2H_2$[76] (e.g., Moses et al. 2011). On highly irradiated giant planets, C and CH are readily produced through photolysis of CO and $CH_4$, and a non-trivial portion of that carbon ends up in $C_3H_2$ and $C_3H_3$. Other than the self reaction of propargyl (a $C_3H_3$ isomer) and reactions with hydrogen, there is very little kinetic information currently available on reactions of $C_3H_3$ and $C_3H_2$ with key hydrocarbon molecules and radicals to form higher-order hydrocarbons (which could eventually form hazes), or with other abundant high-altitude species, such as O, C, N, OH, and $H_2O$.

The diversity of exoplanetary conditions and the much-needed versatility of our interpretation tools both stress the importance of model evaluation as well as the need for mechanism reduction under situations where a chemical models is coupled with models describing complex physical processes where computational expense becomes a critical issue. It is important to determine the predictability of any model which incorporates a parametric-based chemical mechanism and therefore to assess the confidence that can be placed in simulation results. Uncertainty analysis allows the calculation of the overall uncertainty of the simulation results based on the best available knowledge of the input parameters, potentially putting error bars on model predictions. Sensitivity analysis can provide the subsequent identification of the most important parameters driving model uncertainty. These methods can help the identification of important parameters that can be targeted by further laboratory studies, and can form a key part of the process of exoplanetary chemical model evaluation and improvement. Mechanism reduction techniques can also identify the core reactions in a large chemical network and the application of reduced mechanisms may speed up the simulations, allowing engineering optimizations[114].

**Photoabsorption Cross-Sections at High Temperature**

For now, observational constraints are only available for exoplanets which are strongly irradiated by their parent star, as current techniques favour the detection and characterization of exoplanets located very close to their parent stars. The planetary atmospheres targeted for spectral characterization with current and forthcoming instruments are therefore hot, with temperatures varying roughly from 500 to 2500 K, and with UV photochemistry playing an important role in their atmospheric disequilibrium chemistry.

Available photoabsorption cross-sections are usually derived from ambient or low temperature measurements. Data corresponding to higher temperatures are in consequence very scarce. When existing at all, they have been usually performed in the frame of combustion studies, on monochromatic beams or restricted wavelength ranges - usually above 190 nm - while the 100-200 nm range is extremely important for atmospheric chemistry. It has been shown that the absorption of the UV flux increases substantially at wavelengths above 160 nm together with the temperature[115,116,117]. Therefore, when using room temperature laboratory data to model hot exoplanetary atmospheres, the photoabsorption of the UV flux and, consequently, the associated photodissociation rates, are likely very underestimated. Campaigns of measurements at the synchrotron BESSY (Berliner Elektronenspeicherring-Gesellschaft für Synchrotronstrahlung, GERMANY) led by LAB (Laboratoire d'Astrophysique de Bordeaux, FRANCE) and LISA (Laboratoire Interuniversitaire des Systèmes Atmosphériques, Paris, FRANCE) were recently performed to measure the photoabsorption cross section of carbon dioxide $CO_2$ up to 800 K. The photoabsorption cross section increases by four orders of magnitude at 200 nm when heating the gas from 300 to 800 K. The important temperature dependency of the photoabsorption cross sections of $CO_2$ has been found to impact significantly the chemical modeling of the well-studied prototypical hot Neptune GJ 436b[118].

Consequently, the lack of photoabsorption cross-sections at high temperature should now be addressed more thoroughly by measuring photoabsorption cross-sections and their temperature dependency up through the VUV wavelength range (115-230 nm) for the suite of important molecules (e.g. $N_2$, $O_2$, $O_3$, $H_2O$, CO, $CO_2$, $CH_4$, $NH_3$, Na, K, Li, Rb, Cs, TiO, VO, HCN, $C_2H_2$, $H_2S$, $PH_3$) of planetary atmospheres.

**Conclusions**

This white paper outlines several areas where laboratory or *ab initio* investigations could lead to significant model improvements, thereby helping to ensure that exoplanet atmospheric models are best prepared for the large quantity of high-quality ground- and space-based observations that will be acquired over the next two decades. Exoplanet spectral and climate models would benefit from dramatically extended pressure broadening parameters in linelist databases, improved collections of dimer and collision-induced absorption opacities over a wide range of temperatures, and the creation and curation of reliable opacity databases for interpreting high spectral resolution observations of hot exoplanet atmospheres. Cloud models can be substantially improved through laboratory investigations of haze and condensate formation, as well as associated optical properties, over a wide range of atmospheric chemical and thermal conditions. Finally, atmospheric photo- and thermo-chemical models could be advanced by investigations of reaction rate constants for key species over a range of conditions relevant to exoplanet atmospheres, as well as studies of gas photo-absorption cross sections at high temperatures. Of course, some of the proposed investigations may not be feasible given current (or planned) laboratory and computational equipment.

In the near future, the NExSS website at www.nexss.info will begin hosting a section of its web site devoted to a forum where investigators with laboratory expertise, or ab initio calculation expertise, can make their capabilities better known. In addition, those working in atmosphere modeling can refine and expand on the suggestions made here, and include new "needs" as the field progresses. We suggest that a joint panel of atmospheric modelers and laboratory investigators could provide insights into the data that are most sorely needed versus what is technically feasible. Through such a dialogue, leading to improvements in the most fundamental inputs to exoplanet atmosphere models, we strive to meet the observational challenges of the next twenty years.

**Acknowledgements**

This document began as collaborative white paper suggested at the inaugural meeting of NASA's Nexus for Exoplanet System Science (NExSS). NExSS a cross-divisional initiative to advance research in areas that support the search for life beyond the solar system, and includes support from NASA's Astrophysics and Planetary Sciences Divisions.13